\documentclass[a4paper,12pt]{article}
\usepackage{cite}
\newcommand{\be}{\begin{equation}}
\newcommand{\ee}{\end{equation}}
\newcommand{\bea}{\begin{eqnarray}}

\newcommand{\eea}{\end{eqnarray}}

\begin{document}
\def\bea{\begin{eqnarray}}

\def\eea{\end{eqnarray}}
\title{\bf {Quasinormal Modes, Reduced Phase Space and  Area Spectrum of Black Holes}}
 \author{M.R. Setare  \footnote{E-mail: rezakord@ipm.ir}
  \\{Physics Dept. Inst. for Studies in Theo. Physics and
Mathematics(IPM)}\\
{P. O. Box 19395-5531, Tehran, IRAN }}
\date{\small{}}

\maketitle

\begin{abstract}
Motivated by the recent interest in quantization of black hole area spectrum, we consider the area spectrum of
 Schwarzschild, BTZ, extremal Reissner-Nordstr\"om, near extremal Schwarzschild-de
Sitter, and Kerr black holes. Based on the proposal by Bekenstein
and others that the black hole area spectrum is discrete and
equally spaced, we implement Kunstatter's method to derive the
area spectrum for these  black holes. We show that although as
Schwarzschild black hole the spectrum is discrete, it is non
equispaced in general. In the other hand the reduced phase space
quantization is another technique which we discuss here. However
there is a discrepancy between the result of the reduced phase
space methodology and quasinormal modes approach for area spectrum
of some black holes.
\end{abstract}



\newpage
\baselineskip=18pt

\section{Introduction}
Hawking was the first to observe that black holes are not
completely black but they emit radiation \cite{haw1}. This
radiation is essentially thermal and the black hole emits field
quanta of all frequencies. Now, may be one can ask this question:
do quantum black holes have a discrete spectrum? Several authors
have raised the possibility that Hawking radiation might in fact
have a discrete spectrum. This idea was first proposed by
Bekenstein in 1974 \cite{bek1}. According to Bekenstein's
proposal, the eigenvalues of the black hole event horizon area
are of the form \be A_n=\alpha  l_{p}^{2} n \hspace{1ex}, \ee
where $\alpha$ is a dimensionless constant, $n$ ranges over
positive integers and, by using gravitational units $G=c=1$,
$l_p=\hbar ^{1/2}$ is the Planck length. The black hole
entropy-law ($S=\frac{A}{4\hbar}$) combined with a statistical
interpertation of the entropy requires that $\exp{A}{4\hbar}$ is
an integer. This, in turn, constraints the spacing constant
parameter $\alpha$ in Eq.(1) to takethe form $\alpha=4\ln b$ for
some integer $b$. Endorsement for Bekenstein's proposal was
provided by the observation that the area of the horizon $A$
behaves, for a slowly changing black hole, as an adiabatic
invariant \cite {bek2}. It is significant that a classical
adiabatic invariant corresponds to a quantum observable with a
discrete spectrum, by virtue of Ehrenfest's
principle.\\
Since the original heuristic arguments of Bekenstein there has
been a substantial amount of work in trying to derive the
spectrum (1) by more rigorous means \cite{bek1,gour4}. Of
particular relevance to the upcoming analysis is a program that
was initiated by Barvinsky and Kunstatter \cite{bk}. Their
methodology is based on expressing the black hole dynamics in
terms of a reduced phase space and then applying an appropriate
process of quantization. One vital assumption was required in
this analysis; namely, the authors assumed that the conjugate to
the mass is periodic over an interval of $\frac{2\pi}{k}$ where
the $k$ is the surface gravity at the horizon. They did, however,
justify this input by way of Euclidean considerations.\\
In the other hand, recently, the quantization of the black hole
area has been considered \cite{hod1, LQG} as a result of the
absorption of a quasi-normal mode
excitation\cite{reggeW,Chandra1}. Any non-dissipative systems has
modes of vibrations, which forming a complete set, and called
normal modes. Each mode having a given real frequency of
oscillation and being independent of any other. The system once
disturbed continues to vibrate in one or several of the normal
modes. On the other hand, when one deals with open dissipative
system, as a black hole, instead of normal modes, one considers
quasi-normal modes for which the frequencies are no longer pure
real, showing that the system is loosing energy.  The possibility
of a connection between the quasinormal frequencies of black
holes and the quantum properties of the entropy spectrum was
first observed by Bekenstein \cite{bek3}, and further developed
by Hod \cite{hod1}. In particular, Hod proposed that the real
part of the quasinormal frequencies, in the infinite damping
limit, might be related via the correspondence principle to the
fundamental
quanta of mass and angular momentum.\\
In this paper we would like to discuss on the discrepancy between
the reduced phase space methodology and QNM approach for area
spectrum of some black holes.
\section{Quasinormal Modes and Area Spectrum}
It can be shown that for any periodic system, there axists an
adiabatic invariant, which can be calculated (up to a constant
shift) as follows \be I\equiv \oint pdq \propto \int {dE\over
\omega(E)}, \label{bohr} \ee where $(p,q)$ are its phase space
variables, $E$ is the energy of system and $\omega (E)$ is
vibrational frequency. Via Bohr-Sommerfeld quantization has an
equally spaced spectrum in the semi-classical (large $n$) limit:
\be I \approx n\hbar. \label{smi} \ee From this viewpoint, the
main problem of black hole quantum mechanics has been to
correctly identify the physically relevant period of vibrational
frequency. Recently Hod \cite{hod1} assumed an equally spaced
area spectrum and used the apparent existence of a unique
quasinormal mode frequency in the large damping limit to uniquely
fix the spacing.\\
In this section we use the observation about quasinormal mode
frequency to drive the general form in the semi-classical limit
of the Bekenstein-Hawking entropy spectrum for Schwarzschild,
BTZ, extremal Reissner-Nordstr\"om, near extremal Schwarzschild-de
Sitter, Kerr black holes and near extremal black branes
\cite{kun,set1,set2,set3, set4, set5}.\\ We start from the
observation that for a Schwarzschild black hole of mass $M$ and
radius $R$, the real part of the quasinormal mode frequency
approaches a fixed non-zero  value in the large damping limit
as\cite{kok1,nol} \be \omega_{R}=\frac{\ln3}{4\pi R}
\label{schqu} \ee Following \cite{hod1, LQG}, Kunstatter in his
recent interesting paper \cite{kun} assume that this classical
frequency plays an important role in the dynamics of the black
hole and is relevant to its quantum properties. He consider
$\omega_{R}$ to be a fundamental vibrational frequency for a
black hole of energy $E=M$. Using Eq.(\ref{bohr}), we are lead to
the following adiabatic invariant: \be I=4\pi\int\frac{dE R
}{\ln3}=\frac{\hbar S_{BH}}{\ln3}+c, \label{entro} \ee where we
have used the fact that $R=2M=2E$ and the definition of the
Bekenstein-Hawking entropy $ S_{BH}=\frac{A}{4\hbar}=\frac{\pi
R^{2}}{4\hbar}$. Eq.(\ref{smi}) then implies that the entropy
spectrum is equally spaced: \be S_{BH}=n \ln3 \label{spec}  \ee
Then, the degeneracy of quantum states is given by: \be
\Omega(E)=\exp(S_{BH})=3^{n} \label{dege} \ee In this case also
we obtain \be \Delta A=4\hbar \ln3 \label{spac1}\ee then the
spacing constant parameter $\alpha$ in Eq.(1) fine as \be\alpha=4
\ln3 \label{alpeq} \ee It has been suggested that the presence of
$\ln3$ implies a change in the loop quantum gravity gauge group
from $SU(2)$ to $SO(3)$ \cite{LQG}. Now we would like to obtain
the area spectrum of non-rotating BTZ black hole. The non rotated
BTZ black hole line element is as as following
\cite{banados1,banados2}\be ds^2 =-(-M+ \frac{r^2}{l^2} )dt^2
+\frac{dr^2}{(-M+ \frac{r^2}{l^2})}+r^{2} d\theta ^2,
\label{metric2}\ee which has an horizon at \be
r_{+}=\sqrt{\frac{M}{\Lambda}}, \label{hori} \ee and is similar
to Schwarzschild black hole with the important difference that it
is not asymptotically flat but it has constant negative
curvature. The quasi-normal frequencies for non-rotating BTZ
black hole have been obtained by Cardoso and Lemos in
\cite{CFT-Correspond} \be \omega=\pm m-2iM^{1/2}(n+1)
\hspace{1cm} n=0,1,2,...\label{quas} \ee where $m$ is the angular
quantum number. Let $\omega=\omega_{R}-i\omega_{I}$, now by
taking $\omega_{R}$ as previous example, we have \be I=\int
\frac{dE}{\omega_{R}}=\int \frac{1 }{\pm m}dM = \frac{M}{\pm
m}+c, \label{adin} \ee where $c$ is a constant, therefore the
mass spectrum is equally spaced
\begin{eqnarray}
&&
M=mn\hbar, \hspace{0.5cm}m\geq 0 \nonumber \\
&&  M= -mn\hbar , \hspace{0.5 cm}m<0 \label{massspec}.
\end{eqnarray} Then for mass spacing we have \be \Delta M=\pm
m\hbar, \label{spacing} \ee which is the fundamental quanta of
black hole mass. As one can see the mass spectrum is equally
spaced only for a fixed $m$. For different $m$ there are
multiplets with different values of spacing which is given by
Eq.(\ref{spacing}). On the other hand, the black hole horizon area
is given by \be A=2\pi r_{+}. \label{area1} \ee Using
Eq.(\ref{hori}) we obtain \be A=2\pi \sqrt{\frac{M}{\Lambda}}.
\label{area2} \ee The Bohr-Sommerfeld quantization law and
Eq.(\ref{adin}) then implies that the area spectrum is as
following \cite{set1}, \be A_{n}=2\pi \sqrt{\frac{nm
\hbar}{\Lambda}} .\label{spec} \ee As one can see although the
mass spectrum is equally spaced but the area spectrum is not
equally spaced.\\In \cite{bir3} Birmingham et al have shown that
the quantum mechanics of the rotating BTZ black hole is
characterized by a Virasoro algebra at infinity. Identifying the
real part of the quasi-normal frequencies with the fundamental
quanta of black hole mass and angular momentum, they found that
an elementary excitation corresponds exactly to a correctly
quantized shift of the Virasoro generator $L_{0}$ or $\bar{L_0}$
in this algebra. Similar to \cite{bir3}we have not found a
quantization of horizon area. The result for the  area of event
horizon is not equally spaced, in contrast with area spectrum of
Schwarzschild black hole.\\Now we would like to obtain the area
and entropy spectrum of extremal Reissner-Nordstr\"om  (RN) black
holes. The RN black hole's (event and inner) horizons are given
in terms of the black hole parameters by \be
r_{\pm}=M\pm\sqrt{M^{2}-Q^{2}}, \label{hor} \ee where $M$ and  $Q$
are respectively mass and charge of black hole. In the extreme
case these two horizons are coincides \be r_{\pm}=M, \hspace{1cm}
M=Q. \label{ext} \ee According a very interesting conclusion
follows \cite{nitz}(see also more recent paper \cite{and})  , the
real part of the quasinormal frequency for extremal RN black holes
coincides with the Schwarzschild value \be
\omega_{R}^{RN}=\frac{\ln 3}{4\pi R_{H}}, \label{quaf} \ee where
\be R_{H}=2M. \label{rfun} \ee
 Now by taking $\omega_{R}^{RN}$ in this context,
we have \be I=\int \frac{dE}{\omega_{R}^{RN}}=\int \frac{4\pi
R_{H}}{\ln 3}dM =\frac{4\pi}{\ln 3}\int 2M dM= \frac{4\pi}{\ln 3}
M^{2}+c, \label{adinv} \ee where $c$ is a constant. In the other
hand, the black hole horizon area is given by \be A=4\pi
r_{+}^{2}. \label{area1} \ee Which using Eq.(\ref{ext}) in
extremal case is as following \be A=4\pi M^{2}. \label{area2} \ee
The Boher-Sommerfeld quantization law and Eq.(\ref{adinv}) then
implies that the area spectrum is equally spaced, \be A_{n}=n
\hbar \ln 3.\label{spec} \ee By another method we can obtain above
result. From Eq.(\ref{area2}) we get \be \Delta A =8\pi M \Delta
M=8\pi M \hbar  \omega_{R}^{RN} \label{delta}\ee where we have
associated the energy spacing with a frequency through $\Delta M
=\Delta E=\hbar\omega_{R}^{RN}$. Now using Eqs.(\ref{quaf},
\ref{rfun})we have \be \Delta A=\hbar \ln3, \label{spacing} \ee
therefore the extremal RN black hole have a discrete spectrum as
\be A_{n}=n\hbar \ln3. \label{spec2} \ee Which is exactly the
result of Eq.(\ref{spec}). Using the definition of the
Bekenstein-Hawking entropy we have \be
S=\frac{A_{n}}{4\hbar}=\frac{n\ln3}{4}. \label{entro} \ee
\\ Now our aim is to obtain the area and entropy
spectrum of near extremal Schwarzschild- de Sitter black holes.
The metric of the Schwarzschild-de Sitter spacetime is as
following \be ds^{2}=-f(r)dt^{2}+f^{-1}(r)dr^{2}+r^{2}d\Omega^{2},
\label{metr} \ee where \be
f(r)=1-\frac{2M}{r}-\frac{r^{2}}{a^{2}}, \label{metel} \ee where
$M$ is the mass of black hole, and $a$ denoting the de Sitter
curvature radius, which related to the cosmological constant by
$a^{2}=\frac{3}{\Lambda}$. The spacetime possesses two horizons,
the black hole horizon is at $r=r_{b}$, and the cosmological
horizon is at $r=r_{c}$, with $r_b< r_c$.  The third zero of
$f(r)$ locates at $r_0=-(r_b+r_c)$. It is useful to express $M$
and $a^{2}$ as function of $r_b$, $r_c$ \be
a^2=r_{b}^{2}+r_br_c+r_{c}^{2} \label{aeq} \ee and \be
2Ma^2=r_br_c(r_b+r_c) \label {meq} \ee
 The surface gravity $k_b$
associated with the black hole horizon, as defined by the
relation $k_b=\frac{1}{2}\frac{df}{dr}|_{r=r_b}$. Easily one can
show \be k_b=\frac{(r_c-r_b)(r_b-r_0)}{2a^{2}r_b}.
 \label{surgr}\ee Let us now specialize to the near extremal SdS
 black hole, which is defined as the spacetime for which the
 cosmological horizon is very close to the black hole horizon.
 According a recent
study \cite{Cardoso}(based on identifying the relevant scattering
potential with that of the Poscl-Teller model \cite{{PT},{FM}})
the quasinormal spectrum of SdS space depends strongly on the
orbital angular momentum of the perturbation field. Moreover, in
the near extremal SdS case, the real part of the frequency, goes
almost linearly with orbital angular momentum. In a more  recent
paper \cite{ab1}, the quasinormal mode spectrum was calculated for
 SdS space by way of the monodromy method
\cite{motl}. When this form of the spectrum is then subjected to
the near extremal  limit, as was done explicitly in \cite{ab1},
there  is absolutely no orbital angular momentum dependence in
evidence. In recent paper by Medved and Martin \cite{med1} a
possible resolution for this discrepancy have provided. According
to this paper, the monodromy-based calculation, although
perfectly valid in the non-degenerate regime, can not necessarily
be extrapolated up to the point of horizon coincidence.\\
The analytical quasinormal mode spectrum for the near extremal SdS
black hole has been derived by Cardoso and Lemos \cite{car} is as
following \be
\omega=k_b[-(n+1/2)i+\sqrt{\frac{v_0}{k_{b}^{2}}-1/4}+o[\Delta]],
\hspace{1cm}n=0,1,2,... \label{quasi} \ee where \be
v_0=k_{b}^{2}l(l+1), \label{scaler} \ee for scalar and
electromagnetic perturbations, and \be v_0=k_{b}^{2}(l+2)(l-1),
\label{gravi} \ee for gravitational perturbations, $l$ is the
angular quantum number. In Eq.(\ref{quasi}) $\Delta$ is squeezing
parameter \cite{med1} which is as following \be
\Delta=\frac{r_c-r_b}{r_b}\ll 1. \label{deleq} \ee  Now by taking
$\omega_{R}$ in this context, we have \be I=\int
\frac{dE}{\omega_{R}}=\int \frac{dM
}{k_b\sqrt{\frac{v_0}{k_{b}^{2}}-1/4}} =
\frac{M}{k_b\sqrt{\frac{v_0}{k_{b}^{2}}-1/4}}+c, \label{adin} \ee
where $c$ is a constant. Boher-Sommerfeld quantization then
implies that the mass spectrum is equally spaced, \be M=n\hbar
k_b\sqrt{\frac{v_0}{k_{b}^{2}}-1/4}. \label{masspec}\ee The use
of Eqs.(\ref{meq},\ref{aeq}) leads us to \be \delta M=\frac{r_b
\Delta r \delta r_{b}}{2a^{2}}=\hbar
k_b\sqrt{\frac{v_0}{k_{b}^{2}}-1/4}. \label{delmas} \ee In the
other hand, the black hole horizon area is given by \be A_b=4\pi
r_{b}^{2}, \label{area} \ee by the variation of the black hole
horizon and use of Eq.(\ref{delmas}) we have \be \delta A_b=8 \pi
r_b \delta r_b=8\pi \frac{2a^2\hbar
k_b\sqrt{\frac{v_0}{k_{b}^{2}}-1/4}}{\Delta r}. \label{delare}
\ee Now by using Eq.(\ref{area}), one can obtain \be \delta
A_b=24 \pi \hbar\sqrt{\frac{v_0}{k_{b}^{2}}-1/4}. \label{delare1}
\ee Similar result for $\delta A_b$ by another method have been
obtained in\cite{abd}. Then we can obtain the quantization of a
near extremal SdS black hole area as \be A_b=24 \pi n
\hbar\sqrt{\frac{v_0}{k_{b}^{2}}-1/4}. \label{areasp} \ee
 Using the definition of the
Bekenstein-Hawking entropy we have \be S=\frac{A_{b}}{4\hbar}=6\pi
n \sqrt{\frac{v_0}{k_{b}^{2}}-1/4}. \label{entro} \ee\\
Now we extend directly the Kunstatter's approach \cite{kun} to
determine mass and area spectrum of Kerr and extreme Kerr black
holes. The metric of a four-dimensional Kerr black hole given in
Boyer-Lindquist coordinates is \be
ds^{2}=-(1-\frac{2Mr}{\Sigma})dt^{2}-\frac{4Mar
\sin^{2}\theta}{\Sigma}dtd\varphi+\frac{\Sigma}{\Delta}dr^{2}+
\Sigma d\theta^{2}+(r^2+a^2+2Ma^2r\sin^{2}\theta) \sin^{2}\theta
d\varphi^{2} \label{met} \ee where \be \Delta=r^2-2Mr+a^2
\label{deleq} \ee \be \Sigma=r^2+a^2\cos\theta \label{sigeq} \ee
and $M$ is the mass of black hole. The roots of $\Delta$ are
given by \be r_{\pm}=M\pm \sqrt{M^{2}-a^{2}} \label{root} \ee
where $r_{+}$ is the radius of the event (outer) black hole
horizon and $r_{-}$ is the radius of the inner black hole horizon.
In addition, we have defined the specific angular momentum as \be
a=\frac{J}{M} \label{aeq} \ee where $J$ is the angular momentum
of the black hole. The Kerr black hole is rotating with angular
velocity \bea
\Omega&=&\frac{a}{r_{+}^{2}+a^{2}}\\
&=&\frac{J}{2M\left(M^{2}+\sqrt{M^{4}-J^{2}}\right)} \label{angve}
\eea which has been evaluated on the event black hole horizon. In
gravitational units, the Kerr black hole horizon area and its
Hawking temperature are given, respectively, by \bea
A&=&4\pi (r_{+}^{2}+a^{2})\\
&=&8\pi\left(M^{2}+\sqrt{M^{4}-J^{2}}\right) \label{area} \eea and
\bea
T_{H}&=&\frac{r_{+}-r_{-}}{A}\\
&=&\frac{\sqrt{M^{4}-J^{2}}}{4\pi
M\left(M^{2}+\sqrt{M^{4}-J^{2}}\right)} \hspace{1ex}.
\label{temeq} \eea
\\
By applying Bohr's correspondence principle, Hod\cite{hod2}
conjectured that the real part of the asymptotic quasinormal
frequencies of Kerr black hole is given by the formula \be
\omega_{R}=T_{H}\ln3+m\Omega \label{quasi}, \ee where $m$ is the
azimuthal eigenvalue of the oscillation. There was compelling
evidence that the conjectured formula (\ref{quasi}) must be wrong.

\par\noindent
A systematic exploration of the behavior of the first few
overtones, i.e., small values of the principal quantum number
$n$, was first accomplished by Onozawa \cite{onoz}. Onozawa used
the Leaver's continued fraction method to carry out the numerical
calculations. Berti and Kokkotas \cite{berti1} confirmed
Onozawa's results and extended them to higher overtones, i.e.,
highly damped QNMs. They found that the formula conjectured by
Hod, i.e., equation (\ref{quasi}), does not seem to provide a
good fit to the asymptotic modes. Furthermore, Berti and Kokkotas
showed that, as the mode order increases, modes having their
orbital angular momentum eigenvalue $l$ and azimuthal eigenvalue
$m$ to satisfy $l=m=2$, are fitted extremely well by the
relation\footnote{A more sophisticated study of highly damped
Kerr QNMs is performed in \cite{berti2}. In this work the authors
provide complementing and clarifying results that were presented
in previous works\cite{onoz,berti1}.} \be \omega=2\Omega +
\mathit{i} 2\pi T_{H} n \hspace{1ex}. \ee where the term $2\pi
T_{H}$ that appears in the imaginary part of the mode frequencies
is the surface gravity of the event horizon of the Kerr black
hole. Therefore, the real part of the asymptotic frequencies
having $l=m=2$ is given by the expression \be \omega_{R}=2\Omega
\hspace{1ex}. \ee\\
The first law of black hole thermodynamics now takes the form \be
dM=\frac{1}{4}T_{H}dA+\Omega dJ \label{kerr1law} \ee and
obviously the corresponding expression for  adiabatic invariant
is now given by the expression \be I= \int \frac{dM-\Omega
dJ}{\omega_{R}} \label{adiabkerr} \hspace{1ex}. \ee The real part
($\omega_{R}$) of the asymptotic (highly damped) quasinormal
frequencies of the Kerr black hole is given by equation
(\ref{quasi1}) and the angular velocity is given by equation
(\ref{angve}).

\par\noindent
Therefore, the adiabatically invariant integral (\ref{adiabkerr})
is written as \be I=\frac{2}{mJ}\int M
\left(M^{2}+\sqrt{M^{4}-J^{2}}\right)\,\, dM - \frac{1}{m}\int dJ
\label{integral} \ee and after integration, we get \be
I=\frac{1}{2mJ}\left[ M^{2}\left(
M^{2}+\sqrt{M^{4}-{J^2}}\right)- J^{2}
ln\left(M^{2}+\sqrt{M^{4}-{J^2}}\right)\right] -\frac{1}{m}J
\label{action1} \hspace{1ex}. \ee By equating expressions
(\ref{smi}) and (\ref{action1}), we get \be
M^{2}\bar{A}-J^{2}ln\bar{A}-c =\left(2mJl^{2}_{p}\right)n
\label{nonalgebraic} \ee where  the parameter $c$ is equal to $2
J^{2}$, the quantity $\bar{A}$ is the reduced horizon area \be
\bar{A}=\frac{A}{8\pi} \ee and the area $A$ of the Kerr black
hole horizon is given by equation (\ref{area}). The solution to
equation (\ref{nonalgebraic}) is the principal branch of Lambert
W-function and thus the area of the Kerr black hole is written
\be A=8\pi\left(-\frac{J^{2}}{M^{2}}\right) W_{0}[z]
\label{kerrsoln1} \ee where the argument of Lambert W-function is
given by \be
z=\left(-\frac{M^{2}}{J^{2}}\right)e^{-2\left(1+\frac{m\hbar}{J}
n\right)} \hspace{1ex}. \ee Since we are only interested in
highly damped quasinormal frequencies, i.e. $n\rightarrow\infty$,
we keep only the first term of the series expansion of the
Lambert W-function and we get \be A=8\pi
e^{-2\left(1+\frac{m\hbar}{J} n\right)} \label{Kerrsoln} \ee It
is obvious that the area spectrum, although discrete, is not
equivalently spaced even to first order.

\par\noindent
Hod studied again analytically the QNMs of Kerr black hole
\cite{hod3} and he concluded that the asymptotic quasinormal
frequencies of Kerr black hole are given by the simpler expression
\be \omega_{R}=m\Omega \label{quasi1} \ee which is obviously in
agreement with the aforesaid numerical results of Berti and
Kokkotas. This classical frequency plays an important role in the
dynamics of the black hole and is relevant to its quantum
properties \cite{{hod1},{LQG}}.

\section{Reduced Phase Space and Area Spectrum}
We now consider a specific reduced phase space model of charged
black hole \cite{bk} (see also \cite{das}). The starting point of
this model is the result of \cite{lou}that the dynamics of static
spherically symmetric charged configuration in any classical
theory of gravity in $d-$spacetime dimensions is governed by an
effective action of the form \be \Gamma=\int dt
(P_{M}\dot{M}c^{2}+P_{Q}\dot{Q}-H(M,Q)), \label{effect}  \ee
where $M$ and $Q$ are the mass and the charge respectively and
$P_{M}$, $P_{Q}$ the corresponding conjugate momenta. The
momentum $P_{M}$ has the interpretation of asymptotic time
difference between the left and right wedges of a Kruskal
diagram. Note that $H$ is independent of $P_{M}$ and $P_{Q}$,
such that from Hamilton's equations, $M$ and $Q$ are constant of
motion. Now, to incorporate thermodynamics of black holes, one
assumes that the conjugate momentum $P_{M}$ is periodic with
period equal to inverse Hawking temperature times $\hbar$. That
is,\be P_{M}\sim P_{M}+\frac{\hbar}{k_{B}T_{H}(M,Q)}.
\label{invers}\ee Similar assumptions were made in the past using
different arguments \cite{kast}. Note that the above
identification implies that the $(M, P_{M})$ phase subspace has a
wedge removed from it, which makes it difficult, if not possible
to quantize on the full phase-space. Thus, one can make a
canonical transformation $(M,Q,P_{M},P_{Q})\rightarrow
(X,Q,\Pi_{X},\Pi_{Q})$, which on the one hand opens up the phase
space, and on the other hand, naturally incorporate the
periodicity Eq.(\ref{invers})\cite{lou} \be
X=\sqrt{\frac{\hbar(S_{BH}-S_{0}(Q))}{\pi k_{B}}}\cos (2\pi
P_{M}k_{B}T_{H}/\hbar) \label{xeq} \ee \be
\Pi_{X}=\sqrt{\frac{\hbar(S_{BH}-S_{0}(Q))}{\pi k_{B}}}\sin (2\pi
P_{M}k_{B}T_{H}/\hbar) \label{pxeq} \ee \be Q=Q \label{qeq}\ee
\be \Pi_{Q}=P_{Q}+\phi P_{M}+S'_{0}P_{M}T_{H}/k_{B}\label{pqeq}
\ee where the entropy at extremality is given by
$S_{0}(Q)=\frac{\pi k_{B}Q^{2}}{\hbar c}$, $^{'}\equiv
\frac{d}{dQ}$ and $\phi$ is the electrostatic potential at the
horizon. The new phase space is $R^{4}$, on which a rigorous
quantization can be performed in a straightforward fashion.
Moreover, as shown in \cite{bk,das} it is straightforward to
calculate the adiabatic invariant for charged black holes in this
parameterization. In particular, Eqs.(\ref{entro},\ref{spec})
generalize to the following invariant: \be
I\equiv\oint\Pi_{X}dX=\frac{\hbar(S_{BH}-S_{0}(Q))}{2\pi
k_{B}}=(2n+1)\hbar \label{invar1} \ee Quantization yields the
following spectra for entropy and charge of the four dimensional
quantum black hole( for more details refer to the \cite{das}): \be
S_{BH}=(2n+p+1)\pi k_{B},\hspace{1cm}n=0,1,2,...\label{entro2} \ee
\be \frac{Q}{e}=m,\hspace{1cm}m=0,\pm1,\pm2,...\label{chsp}  \ee
\be p\equiv\frac{Q^{2}}{\hbar c}=m^{2}\frac{e^{2}}{\hbar c}
\label{chsp1}\ee where consistency requires $p$ to be a
non-negative integer.\\Recently the above formalism was extended
to rotating (uncharged) black holes in $4$ spacetime dimensions
\cite{gour}\footnote{In recent paper \cite{dap} the authors have
extended the results obtained for a charged black hole in an
asymptotically flat spacetime to the scenario with non vanishing
negative cosmological constant.}. The authors showed that the
corresponding area spectrum is also discrete and equispaced: \be
A=8\hbar\pi(n+m+1/2) \ee where $n$ and $m$ are non-negative
integers, while $n$ signifies the departure of the black hole
from extremality, $m$ measures the classical angular momentum of
the black hole. More recently, Das et al \cite{das3}, extend
their formalism to include BTZ black holes as well as $5$ and
higher dimensional Myeres-Perry type rotating black holes with
multiple angular momenta parameters. They show that while the
area spectrum is discrete in each case in general it is not
equispaced.

\section{Conclusions}
Bekenestein's  idea for quantizing a black hole is based on the
fact that its horizon area, in the nonextremal case, behaves as a
classical adiabatic invariant.  Discrete spectra arise in quantum
mechanics in the presence of a periodicity  in the classical
system which in turn leads to the existence of an adiabatic
invariant  or action variable. Boher-Somerfeld quantization
implies that this adiabatic invariant has an equally spaced
spectrum in the semi-classical limit. In this article we have
reviewed some recent study in black hole area spectrum, including
Kunstatter's approach, and the reduced phase space methodology.
We have extended the Kunstatter's approach for Schwarzschild
black hole to another black holes including, BTZ, extremal
Reissner-Nordstr\"om, near extremal Schwarzschild-de Sitter, and
Kerr black holes, and have shown that although as Schwarzschild
black hole the spectrum is discrete, it is non equispaced in
general. We have showen that while the area spectrum is discrete
in each case in general it is not equispaced. The result for the
area of event horizon in BTZ black hole case is as: $A_{n}=2\pi
\sqrt{\frac{nm \hbar}{\Lambda}}$ which is not equally spaced.
Then we have found that the quantum area for extremal RN black
hole in four dimensional spacetime, is as: $\Delta A=\hbar \ln3$.
Although the real parts of highly damped quasi-normal modes for
schwarzschild and extremal RN black hole is equal \cite{and}
$\omega_{R}=\frac{\ln3}{8\pi M }$, as one can see for example in
\cite{{kun},{bir},{and}}, $\Delta A=4\hbar \ln3$ for
schwarzschild black hole. Therefore in contrast with claim of
\cite{and}$, \Delta A $ is not universal for all black holes.
Also we have found that the quantum area for near extremal SdS
black hole in four dimensional spacetime is as: $\Delta A=24 \pi
\hbar\sqrt{\frac{v_0}{k_{b}^{2}}-1/4}$. Also Abdalla et al \cite
{abd} have shown that the results for spacing of the area
spectrum for near extreme Kerr  black holes differ from that for
schwarzschild, as well as for non-extreme Kerr black holes. Such
a difference for problem under consideration in this letter also
in \cite{abd} as the authors have been mentioned may be justified
due to the quite different nature of the asymptotic quasi-normal
mode spectrum of the near extreme black hole. Finally we have
showen that the area spectrum for Kerr black hole is as: $A=8\pi
e^{-2\left(1+\frac{m\hbar}{J} n\right)}$, it is obvious that the
area spectrum, although discrete, is not equivalently spaced.\\
In the other hand the reduced phase space quantization is another
technique which we have discussed here. However there is a
discrepancy between the result of the reduced phase space
methodology and QNM approach for area spectrum of some black
holes. As for why there is an apparent discrepancy between the
QNM  calculations and that of reduced phase space methodology,
there is no easy answer. It is not clear what role (if any)
quasinormal spectrum plays in quantum theory of gravity (which as
of now is nonexistent). I see only two useful applications of
quasinormal spectrum: classical general relativity and AdS/CFT
correspondence \cite{AdS}, where QNM appear as poles of the
retarded correlations functions in the quantum field theory dual
to a particular gravitational background \cite{stari1}.

\end{document}